\documentclass[conference]{IEEEtran}
\IEEEoverridecommandlockouts
\usepackage{cite}
\usepackage{amsfonts}
\usepackage{graphicx}
\usepackage{textcomp}
\usepackage{xcolor}
\usepackage[colorinlistoftodos]{todonotes}

\usepackage{algorithm}      % For the algorithm float environment
\usepackage{algpseudocode}  % Alternative to algorithmic, but not used in your snippet
\usepackage{algorithmicx}   % Base package for algorithm environments
\usepackage{amsmath}        % For math formatting (\( \), etc.)
\usepackage{amssymb}        % For math symbols (optional but commonly used)
\usepackage{float}          % For using [H] specifier (forces placement)
\usepackage{setspace}       % (optional) For spacing inside the algorithm
\usepackage{mathtools,bbm}      % (optional) Enhances amsmath
\usepackage{caption}        % (optional) For customizing captions
\usepackage{tikz}           % (optional) If you're visualizing assignment graphs
\usepackage{amsmath}

\usepackage{hyperref}      % Optional but improves clickable links

\begin{document}

\title{A Virtual Quantum Network Prototype for Open Access
\thanks{Source code for the prototype is available at: \url{https://github.com/rajmadhu0406/Virtual-Quantum-Network-VQN}. Access to the local server requires setup and authentication. The system is currently in internal testing and not publicly accessible. Interested users may contact the authors via email to request access.}
}

\author{
\IEEEauthorblockN {
Raj Kamleshkumar Madhu$^{\dag,*}$, Visuttha Manthamkarn$^{\$,*}$,  Zheshen Zhang$^{\$}$, and Jianqing~Liu$^{\dag}$
}
\IEEEauthorblockA {
  $^{\dag}$Department of CS, NC State University, Raleigh, USA 27606. \\
  $^{\$}$Department of EECS, University of Michigan, Ann Arbor, USA 48109. \\
  $*$ co-first authors
}
\IEEEauthorblockA {
    $^{\dag}$rmadhu3@ncsu.edu; jliu96@ncsu.edu.
    $^{\$}$visuttha@umich.edu; zszh@umich.edu.
}}

\maketitle

%--------------------------------------ABSTRACT-------------------------------------%

\begin{abstract}
The rise of quantum networks has revolutionized domains such as communication, sensing, and cybersecurity. Despite this progress, current quantum network systems remain limited in scale, are highly application-specific (e.g., for quantum key distribution), and lack a clear road map for global expansion. These limitations are largely driven by a shortage of skilled professionals, limited accessibility to quantum infrastructure, and the high complexity and cost associated with building and operating quantum hardware. 
To address these challenges, this paper proposes an open-access software-based quantum network virtualization platform designed to facilitate scalable and remote interaction with quantum hardware. The system is built around a cloud application that virtualizes the core hardware components of a lab-scale quantum network testbed, including the time tagger and optical switch, enabling users to perform coincidence counts of the photon entanglements while ensuring fair resource allocation. The fairness is ensured by employing the Hungarian Algorithm to allocate nearly equal effective entanglement rates among users. We provide implementation details and performance analysis from the perspectives of hardware, software, and cloud platform, which demonstrates the functionality and efficiency of the developed prototype.
%This algorithm was tested in simulation by generating user requests based on a Poisson process. Simulation results demonstrate the algorithm’s ability to maintain high load balancing and fairness, even under contention for limited resources.
%Currently, the operational platform handles user requests using a First-Come, First-Served (FCFS) approach. However, the system is designed to be scalable and extensible to support long-distance quantum networking experiments in the future, making it a flexible foundation for advancing global quantum research and education.

\end{abstract}

\begin{IEEEkeywords}
Quantum Networks, Virtualization, Resource Allocation
\end{IEEEkeywords}

%--------------------------------------INTRODUCTION-------------------------------------%

\section{Introduction}
Network virtualization refers to abstracting network resources that were traditionally delivered in hardware using software-based configurations. The virtualization technique is the key to rapid and cost-effective scaling of cyber infrastructures. It enables flexible provision of physical resources to a much broader user base. Network virtualization is widely used in public clouds like AWS, Azure and GCP. While classical network virtualization has been well-established, the shared access to public quantum systems, especially quantum entanglement distribution networks, remains in its early stages of development. The IBM Quantum Platform is the world’s first quantum computing cloud platform which offers quantum processing units. When it comes to shared public access to quantum networks, there are very few ongoing efforts. The city-scale quantum network in Bristol \cite{ATrustedNode}, Illinois Express Quantum Network \cite{IEQNET}, and the quantum key distribution (QKD) network in Tokyo \cite{QKDTokyo}) are a few notable quantum networks with network virtualization, but they are specialized, proprietary, and not publicly accessible. A complete survey of existing quantum network testbeds can be found in \cite{liu2024road}.

A quantum network is a communication system that transfers quantum information, a.k.a. quantum bits or qubits, from one node to another, enabling new sensing, computing, and security capacities that are impossible with classical networks. Despite that distributing quantum information does not require genuine quantum resources like entanglement, a quantum network capable of distributing entanglement has been highly sought after because of its support for many novel applications such as distributed quantum sensing and decision making \cite{gu2024fendi,yu2022topology}. However, most of the existing quantum entanglement distribution networks are local, application specific, and not available to public. One significant attempt to build a public quantum entanglement distribution network was done by the University of Illinois \cite{PQN}. Unlike previous quantum networks limited to scientists, this network allows the public to interact with entangled photons and perform real quantum measurements by visiting the Urbana Free library. While this is a valuable step toward public quantum networks, it has several limitations: users must be physically present, it is application-specific, and it supports only one user at a time. To effectively engage a diverse range of scientific communities and accelerate quantum network research, it is crucial to abstract the hardware complexities while providing users with open-access, software-reconfigurable quantum networks.

The motivation behind this paper stems from the need to promote research in quantum technology and democratize access to quantum networks. Traditional quantum systems often dedicate entire hardware setups to single tenant, leading to inefficiencies and high costs. This paper presents a technology called virtual quantum network (VQN), which concerns the abstraction of the daunting hardware operations into more friendly and controllable software functions, thereby hiding hardware complexities and enabling easier access. By further placing the abstracted functions in the internet cloud, VQN permits many users to control quantum resources remotely and concurrently. The benefits of VQN are many-fold. It is designed without assuming users' prior knowledge in operating complex quantum hardware, ultimately lowering the entry barrier for educators and students by simplifying the operation of complex quantum systems. This makes VQN a great tool for training and education. %Ultimately, the research will pave the way for flexible and scalable quantum network infrastructures, empowering a wide range of users to contribute to the field without requiring extensive technical expertise. 
In addition, VQN enhances the network capacity by virtualizing quantum resources for parallel access. The resources can be entangled photons of different wavelengths or paths/channels for entanglement routing. VQN can allocate users with orthogonal resources for non-interfering access, ultimately realizing the full capacity of the physical quantum network. %Through virtualization, the VQN achieves substantially higher network capacity compared to its physical quantum network counterpart. This approach not only lowers the barriers to entry for quantum research but also supports educational innovation and expands access to quantum training for a broader audience. 

%------------------------------------QUANTUM HARDWARE-------------------------------------%

\section{Quantum Entanglement Distribution Network}

\begin{figure}[tbp]
\includegraphics[width=\linewidth]{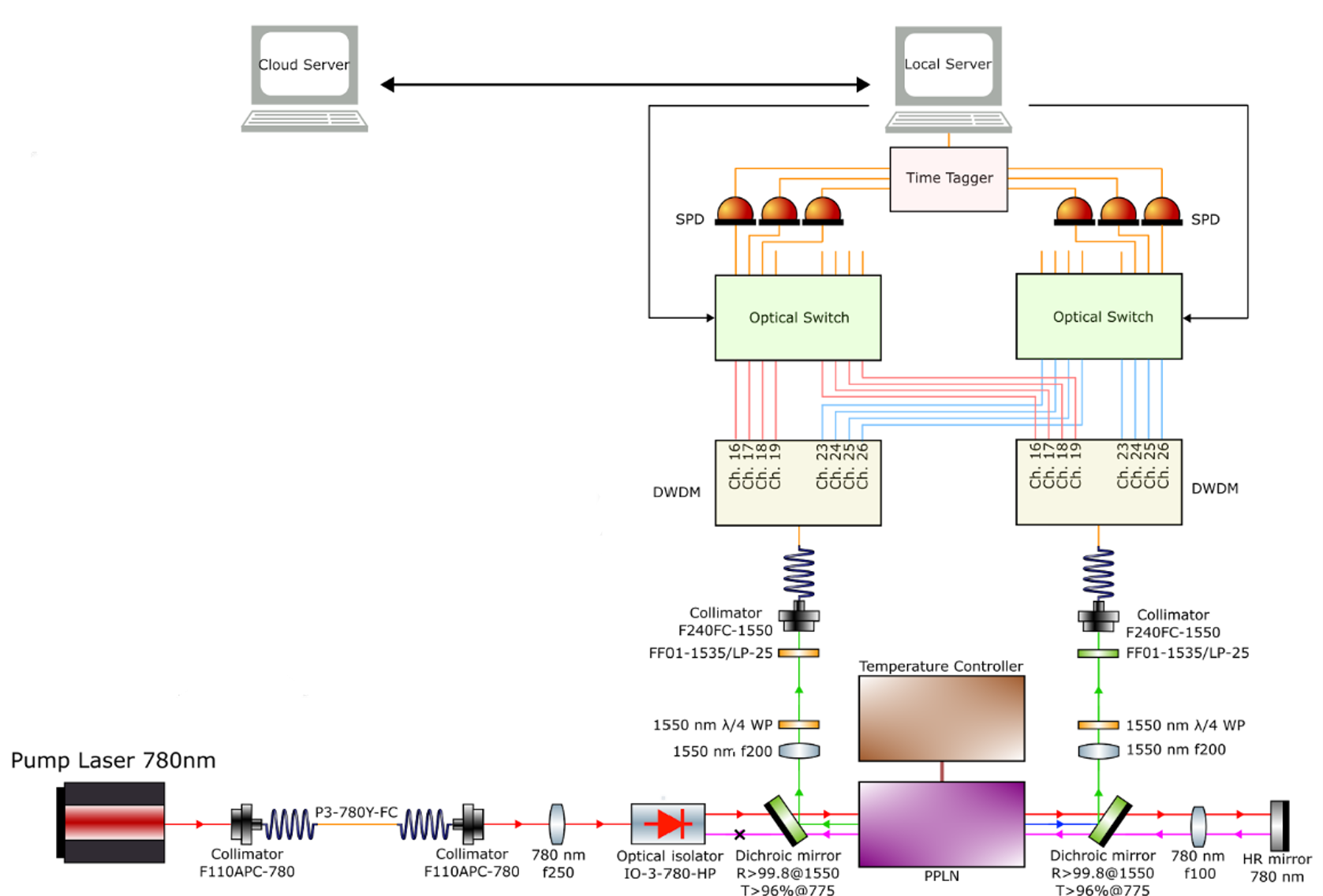}
\caption{Schematic of Hardware Setup}
\label{fig:hardware}
\end{figure}

\begin{figure}[tbp]
\includegraphics[width=\linewidth]{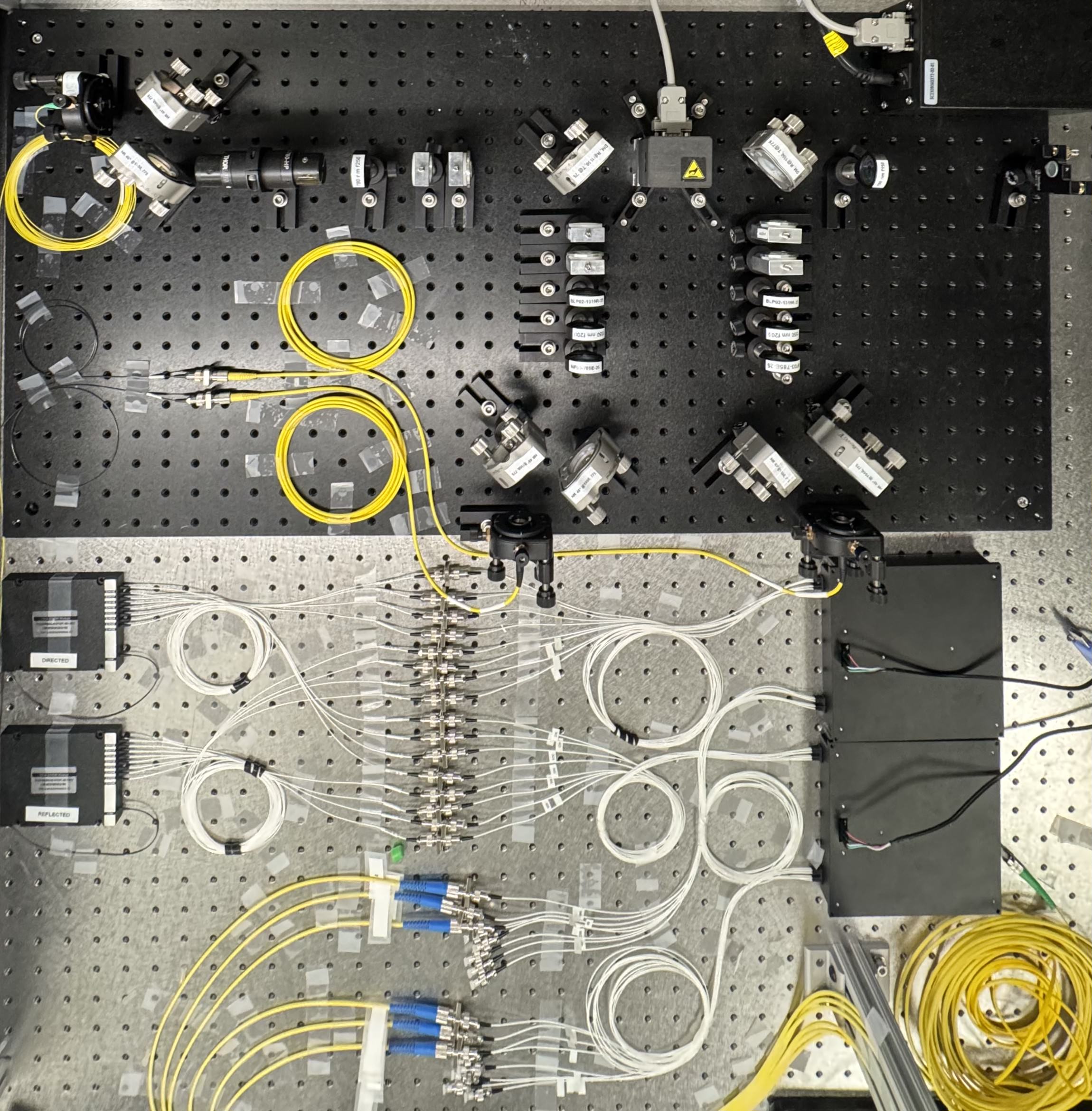}
\caption{Quantum network testbed at the University of Michigan.}
\label{fig:testbed}
\end{figure}

The hardware setup for entanglement generation is illustrated in Figure~\ref{fig:hardware}, while the full testbed configuration, shown in Figure~\ref{fig:testbed}, is located at the University of Michigan, Ann Arbor, in the United States. The experiment begins with a continuous-wave pump laser operating at 780nm, which emits high-energy photons used to drive the entanglement process. These photons are carefully guided and focused using optical elements such as lenses and collimators to ensure high-precision spatial alignment and polarization control—critical parameters for efficient nonlinear interaction in the subsequent stage.

The aligned pump beam is focused into a Periodically Poled Lithium Niobate (PPLN) crystal, a nonlinear optical material configured for type-0 phase matching. Inside the crystal, spontaneous parametric down-conversion (SPDC) occurs: each high-energy pump photon is probabilistically converted into a pair of lower-energy, time-energy entangled photons—designated as the signal (shorter wavelength) and idler (longer wavelength). This process conserves both energy and momentum, enabling deterministic relationships between the generated photon pairs.

The entangled photon pairs, which are broadband and centered around 1560nm, are coupled into optical fibers and spectrally filtered using Dense Wavelength-Division Multiplexing (DWDM). In our system, signal photons are routed into DWDM channels 23–26, while the corresponding idler photons are routed into channels 16–19, following the ITU (International Telecommunication Union) grid standard. This frequency assignment guarantees that the sum of the signal and idler frequencies matches the pump frequency, ensuring energy conservation. For example, idler channel 19 (1562.23nm) is paired with signal channel 23 (1558.98nm).

The standardized ITU channel mapping allows seamless integration with existing telecom infrastructure and promotes compatibility with future quantum networking equipment. Each signal-idler pair is routed through MEMS-based optical switches, which serve as the distribution layer. The switches de-multiplex the DWDM signals and selectively direct the photons to their respective detection paths.

Detection is performed using superconducting nanowire single-photon detectors (SNSPDs), known for their high efficiency, low dark count rates, and picosecond timing resolution. The current hardware configuration includes six SNSPDs, supporting three simultaneous entangled channel pairs at any given time.

All detectors are connected to a time tagger—a high-precision device that records the arrival times of photon detection events. These timestamps are crucial for identifying coincidence events, which indicate successful entanglement. The time tagger is interfaced with a local server that exposes an API for programmatic access to functions such as photon counting and coincidence analysis.

A cloud-based control application provides a virtualized interface for remote users, enabling them to submit measurement requests and interact with the quantum hardware in real time. Upon receiving a request, the cloud software dynamically allocates available entangled channel pairs and communicates with the local server to execute the measurement protocol. This virtualized and multiplexed system architecture supports concurrent access by multiple users while maintaining measurement fidelity.

%-------------------HARDWARE SCHEMATIC IMAGE --------------------%
%------------------------------------------------------%

%-------------------HARDWARE IMAGE --------------------%
%------------------------------------------------------%

% %-------------------Components Table--------------------%

% \begin{table}[tbp]
% \caption{Components used in Hardware setup}\label{tab:components}
% \begin{center}
% \begin{tabular}{| c | c |}
% \hline
% \textbf{Componet Name} & \textbf{count} \\
% \hline

% \rule{0pt}{3ex}

%  Pump Laser 780 nm & 1 \\
%  Collimator 780 nm & 2 \\
%  Collimator 1550 nm & 2 \\
%  Single Mode (SM) Fiber 780 nm & 1 \\
%  Single Mode (SM) Fiber 1550 nm & 2 \\
%  Optical Isolator 780 nm & 1 \\
%  Dichroic Mirror & 2 \\
%  Periodically Poled Lithium Niobate (PPLN) & 1 \\
%  Temperature Controller & 1 \\
%  Longpass Filter 1535 nm & 2 \\
%  Lens 780 nm f250 & 1 \\
%  Lens 780 nm f100 & 1 \\
%  Lens 780 nm f200 & 2 \\
%  Quarter Wave Plate (QWP) 1550 nm & 2 \\
%  Long Pass Filter (LP) 1550 nm & 2 \\
%  Plano Mirror 780 nm & 1 \\
%  Dense Wavelength Division Multiplexing (DWDM) & 2 \\
%  Optical Switch & 2 \\
%  Single Photon Detectors (SPD) & 6 \\
%  Time Tagger & 1 \\

% \hline 

% \end{tabular}
% \end{center}
% \end{table}

% %---------------------------------------------------%

%--------------------------------------SOFTWARE-------------------------------------%

\section{Virtual Functions in Cloud}

%------------------- USER REQUEST FLOW IMAGE --------------------%
\begin{figure}[tbp]
\includegraphics[width=\linewidth]{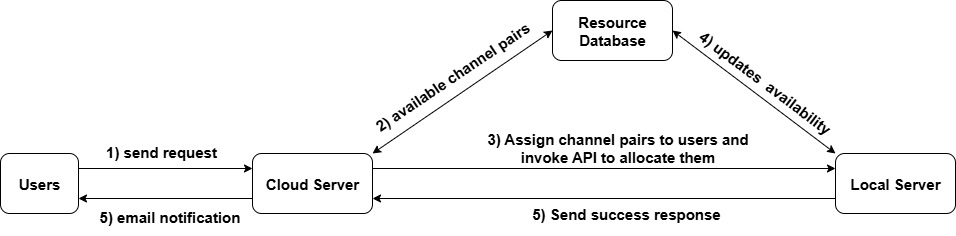}
\caption{User Request Flow.}
\label{fig:user_request}
\end{figure}
%------------------------------------------------------%

Building on the hardware setup described above, the cloud application serves as the central interface that enables remote access to the quantum system. Designed with a modular architecture, the application comprises two primary components: the frontend and the backend. Together, these components manage user interactions and coordinate with the local server to process measurement requests efficiently while maintaining security and fairness in resource usage. The cloud application is developed using FastAPI, MySQL, Redis, Kubernetes and AWS. The backend is integrated with two Redis Pub/Sub channels: the User Request channel, which manages incoming requests for channel pairs from users, and the Response channel, which handles the delivery of resource assignment updates to those users. Similarly, the MySQL server maintains two key databases: the Resource database, which records the current availability of entangled channel pairs within the system, and the Request database, which keeps track of all user requests and their statuses.

Before users can submit measurement requests, they are required to undergo authentication to ensure that only authorized individuals can access the quantum hardware. To request access and experiment on the platform, users are required to contact the authors via email. Once authenticated, users must request access to an available entangled channel pair—established through the hardware setup—before performing any measurements. The user request flow is visualized in  Figure \ref{fig:user_request}. When users submit channel pair requests through the application, the request is simultaneously stored in the Request database with status “processing” and added to the Redis User Request queue. The user immediately receives a confirmation message in the web interface, indicating that their request is being processed. Behind the scenes, a dedicated resource allocation service continuously monitors the Redis queue and checks the Resource database to determine the availability of channel pairs. When resources become available, the service allocates an appropriate channel pair to the user. At present, the system’s resource (channel pair) allocation algorithm is only tested in simulation and currently the application employs a First-Come, First-Served (FCFS) scheduling approach to manage access to the limited number of channel pairs. The application communicates the allocation to the local server, which interfaces directly with the optical hardware and updates the Resource database to reflect the current allocation status. Once the assignment is successfully made, the request status is updated to “processed,” and a response is published to the Response Redis Pub/Sub. The notification service, subscribed to this channel, automatically notifies the user through AWS Simple Email Service (SES) and marks the request as “completed.”
After receiving a channel pair, users can submit specific measurement requests through the cloud application. These requests are forwarded to the local server, which interacts with the time tagger and optical switch—outlined in the hardware setup, to carry out the measurements and return the results to the user. Once the user completes their measurements, the channel pair is released back into the system, making it available for allocation to other users. The entire cloud architecture is designed with loose coupling between services, allowing for independent scaling of each component and ensuring efficient handling of multiple concurrent users. This virtualized and scalable design enables seamless remote access to the quantum hardware while preserving system responsiveness and fairness.

%-------------------------------MEASUREMENT FUNCTIONS--------------------------------%

\subsection{Measurement Functions}

In the current setup, the measurement functions provided to users are as follows:

\subsubsection{Count Rate}

This function measures the average photon detection rate on specified channels. It is implemented using the time tagger and reports the number of photon detection events per second for each selected input channel.

\subsubsection{Counter}

The Counter function allows users to analyze the temporal distribution of photon arrivals. Users specify two input parameters: the bin width (in picoseconds) and the number of bins. The time tagger returns the number of photon events detected within each bin, effectively generating a time-resolved histogram of detection events.

\subsubsection{Counter Graph}

This feature is provided by the cloud application to visualize the histogram data generated by the Counter function. It enables users to interpret temporal detection patterns through an intuitive graphical interface.

% See \textbf{Figure\ref{fig:counter_graph}}.

% %-------------------Counter Graph IMAGE --------------------%
% \begin{figure}[tbp]
% \includegraphics[width=\linewidth]{figs/counter_graph.png}
% \caption{Graph representing Counter Data.}
% \label{fig:counter_graph}
% \end{figure}
% %------------------------------------------------------------%

\subsubsection{Coincidence Count}

The coincidence count function is designed to measure the number of entangled photon pairs generated per second, providing insight into the performance and stability of the quantum source. 
Two photons are considered part of an entangled pair when they are detected within a narrow temporal separation (e.g., 10\,ps), which reflects the strong timing correlation expected from entanglement. In practice, a broader coincidence window (e.g., 500\,ps) is used during post-processing to count these temporally correlated detection events.
By identifying detection events that fall within this window, the system records coincidence counts that serve as a proxy for entangled photon pair generation. 
The function outputs several key metrics: \textit{channel rates}, which represent the individual photon detection rates for each channel; the \textit{coincidence rate}, indicating how frequently entangled photon pairs are detected; and \textit{accidental coincidences}, which refer to uncorrelated photons that fall within the window purely by chance.
Finally, the \textit{coincidence-to-accidental ratio (CAR)} compares the number of true coincidences to accidental ones, providing a useful metric for evaluating entanglement quality.

% \begin{itemize}
%     \item {Channel Rate}: The photon detection rates of both channels in the channel pair.

%     \item {Coincident Rate}: The rate at which the entangled photons are generated at a specific channel pair.

%     \item {Accidental Coincidence Rate}: Even when photons are detected in the same coincidence time window, they might not be entangled. These are called accidental coincidence and they are calculated from the average coincidence count outside the coincidence window.

%     \item {Coincidence-to-accidental Ratio (CAR) Ratio}: It is the ratio of the coincidence count to the accidental coincidence count.
% \end{itemize}

%----------------------------------RESOURCE ALLOCATION--------------------------------%

\section{Resource Allocation}
Entangled qubits is a critical resource in any quantum network, necessitating fair and efficient allocation. Different quantum networks come with varying constraints, driving the need for custom resource allocation algorithms. For example, some researchers proposed utility functions that balance both entanglement fidelity and rate \cite{QNUM}, while others leveraged weighted round-robin \cite{RL-DQC-1} or mixed-integer linear programming \cite{RM-CS-DCQ-2} approaches to optimize fair entanglement distribution. 

In our current testbed setup, we have six single-photon detectors, which allow only three pairs of channels to be measured concurrently. This limitation creates a resource management challenge: how to fairly assign these finite, high-value channel pairs among competing users? Each channel pair has varying entanglement distribution rates and if multiple users request access simultaneously, some may end up with higher-rate channels while others receive lower-quality ones. To maintain an equitable level of service, we require a resource allocation mechanism that ensures fair distribution of these pair of channels. 

\subsection{Quality of Service}

% Session based, use summation , R use notation, summation of total session (denominator)   R(t) fair RA

To measure users' satisfaction level, we define a Quality of Service (QoS) metric that captures both the resource usage and the waiting time experienced by users. The QoS is calculated for each user session and is mathematically expressed as:

% \begin{equation}
% \label{eq:qos}
%     QoS = \frac{Total \ Conincidence \ Count}{Total \ Usage \ Time \ + \ Total \ Wait \ Time}
% \end{equation}\\

\begin{equation}
\label{eq:qos}
    QoS_i = \frac{\int_{0}^{T} \mathbbm{1}_w(i;t) \cdot R_w(t) \,dt}{T}, \quad \forall w \in \overrightarrow{W}
\end{equation}

The QoS for user $i$ is calculated as the ratio of the total number of entangled photon pairs that the user previously received in its session to the total time invested by the user. Specifically, \( R_w (t) \) represents the real-time coincidence rate of a given pair of entanglements $w$. The indicator function $\mathbbm{1}_w(i;t) = 1$ if entanglement pair $w$ is assigned to user $i$ at time $t$; and 0 otherwise. The user $i$ stays in the service for a total time $T$, which consists of waiting time and service time. A user can be a new user or returning one, so the selection of $T$ may have to take into account the service history. In summary, Eq.(\ref{eq:qos}) quantifies the effective entanglement rate in a given session of time $T$; and a higher QoS indicates more efficient use of network resources.
%This ratio serves as an effective average rate for each user, indicating how many entangled pairs are accessed per unit of total time invested (both waiting and using the channel). A higher QoS value implies that the user experiences more entangled photon detections while minimizing their overall delay. Conversely, a lower QoS indicates that a user waited longer or received fewer entangled pairs during their channel usage. By measuring how each user’s entanglement intake stacks up against their time spent waiting and using the resource, the algorithm can systematically align with fairness goals, ensuring a more equitable experience for all participants.

%-----------------------------Proportional Fair Utility Function-------------------------%

\subsection{Fair Resource Allocation}
Now that we have a way to measure QoS, we further impose a fairness function on the QoS metric so that users can obtain a fair share of network resources. In this paper, we adopt the principle of proportional fairness for both efficiency and fairness. The idea is to proportionally allocate entangled photons in different channel pairs (thus different entanglement rate) by prioritizing users with lower QoS, while preserving overall fairness in the system. The utility function, when channel pair $w$ is assigned to user $i$ is given by $U_i(t) = \log\left(1 + \frac{R_w (t)}{QoS_i}\right)$, where \( R_w(t)\) represents the coincidence rate of the channel pair \(w\) at the current time instance $t$. This rate is time-dependent and determined by various hardware parameters such as pump power, alignment efficiency, and thermal stability, as described in Section II. The term $\frac{R_w (t)}{QoS_i}$ accounts for the history of user $i$'s QoS to determine the rate assignment at the current time instance. It thus gives rise to the proportional allocation of entanglement resources to users according to their historical average rates, achieving fairness and efficiency.  %Consequently, \( R_w(t)\) captures the real-time entanglement rate of a channel pair $w$, allowing the utility function to adapt to dynamic system conditions. Meanwhile, \( QoS_i \) reflects quality of service experienced by user \(i\) in their current session. This proportional fairness approach is a well-established concept in wireless scheduling and networking, proven to prevent resource monopolization and ensure equitable distribution among users.

%For quantum resource allocation, proportional fairness offers several advantages. First, it ensures fair treatment of under-served users by sharply increasing utility for those with low QoS, preventing any single user from monopolizing limited quantum channels. Second, it is adaptable, allowing adjustments to accommodate evolving hardware constraints (i.e., fidelity), while maintaining equitable distributions. Simpler alternatives, like maximizing total entanglement or using round-robin scheduling, often fail to address under-served users or compromise system fairness. To fairly allocate channel pairs, we assign the channel pairs to users such that the sum of the utility function is maximized. 

%-----------------------------hungarian matching algorithm-------------------------%
\subsection{Hungarian Matching Algorithm}

Now that we have established our utility function, the next step is to allocate resources to users in a way that maximizes the sum of the utility function. This is essentially an assignment matching problem, where the goal is to pair resources with users to achieve the highest possible total utility. To solve this, we employ the Hungarian Matching Algorithm, which is designed to find the optimal assignment between resources and users while maximizing the sum of the utility function. The Hungarian algorithm is particularly effective in bipartite graph scenarios, such as assigning jobs to workers or resources to users, with the aim of minimizing cost or maximizing payoff. The algorithm takes a cost matrix $C$ of size $W \times N$ where N is the number of users' requests and W is the number of available channel pairs. ${C_{w,i}}$ represents the value of proportional fair utility function for assigning channel pair $w$ to user $i$. The algorithm returns the optimal assignment of channel pairs to users such that the sum of the utility function is maximized. It solves the assignment problem in polynomial time $\mathcal{O}(N^3)$.

% %%%%------ALGORITHM START-----%%%%%

% \begin{algorithm}[H]

% \caption{Hungarian Matching Algorithm}
% \label{alg:hungarian}

% \SetAlgoLined
% \BlankLine

% \KwData{\\ \(n \times m\) cost matrix \(C\) where n is the number of users and m is the available channel links. \\ \(C_{ij}\) is the value of utility function for assigning channel link j to user i.}

% \BlankLine

% \KwResult{\\ Optimal assignment of channel pairs to users such that the sum of utility function is maximized.}
% \BlankLine

% \textbf{Pseudocode: } \\

% Make \(C\) a square matrix by adding dummy rows/columns with zero cost\;

% \BlankLine

% \For {each row \(i\)}{
%   Subtract \(\min(C_{i,*})\) from each element in row \(i\)\;
% }

% \BlankLine

% \For{each column \(j\)}{
%   Subtract \(\min(C_{*,j})\) from each element in column \(j\)\;
% }

% \BlankLine

% \While{optimal assignment is not found}{

% \BlankLine

% Cover all zeros with a minimum number of horizontal and vertical lines\;

% \BlankLine

% \If{number of covering lines equals matrix size}{
%   \Return return matching by assigning one zero per row and column (no duplicates)\;
% }
% \Else{
% Find the smallest uncovered value \(k\)\;
% Subtract \(k\) from all uncovered elements\;
% Add \(k\) to all elements at intersections of covering lines (covered twice)\;
% }

% \BlankLine

% }

% \end{algorithm}

% %%%%------ALGORITHM END-----%%%%%

\section{System Evaluation}
In this section, we present the setups for system evaluation and the results from the hardware, software, and system perspectives. 
\subsection{Hardware Performance}
To evaluate the performance of the entanglement source and distribution system, we conducted measurements using a 10\,mW continuous-wave pump laser operating at 780\,nm. The system was configured to simultaneously support three entangled channel pairs: ITU channels Ch26–Ch16 (1556.55\,nm – 1564.68\,nm), Ch25–Ch17 (1557.36\,nm – 1563.86\,nm), and Ch24–Ch18 (1558.17\,nm – 1563.05\,nm). These correspond to signal-idler photon pairs centered around 1560\,nm, which were routed through DWDM filters and detected using six SNSPDs.

Each detector pair was connected to a Time Tagger 20 (Swabian Instruments), which recorded photon arrival times with 1\,picosecond digital resolution. Using these timestamps, we constructed coincidence histograms and evaluated temporal correlations between the signal and idler channels.

\begin{figure}[tbp]
\includegraphics[width=\linewidth]{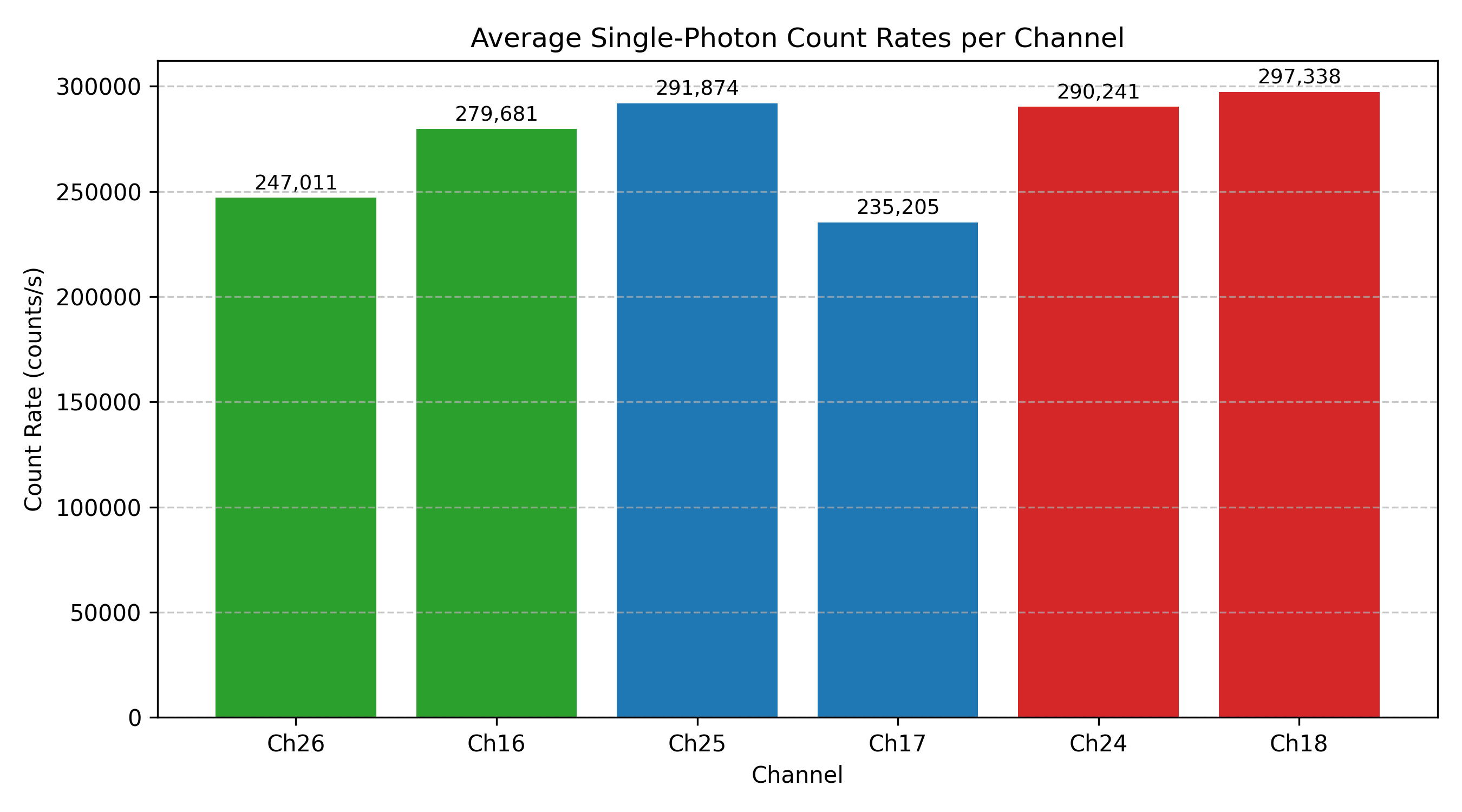}
\caption{Average photon count rates for each DWDM channel.}
\label{fig:avg_count_rate}
\end{figure}

The average photon count rates are shown in Figure~\ref{fig:avg_count_rate}, with channels grouped by entangled signal-idler pairs. All channels exhibit count rates ranging from approximately 230\,k to 300\,k counts per second. The variations are likely due to filter insertion loss, fiber attenuation, polarization drift, or differences in detector efficiency.

\begin{figure}[tbp]
\includegraphics[width=\linewidth]{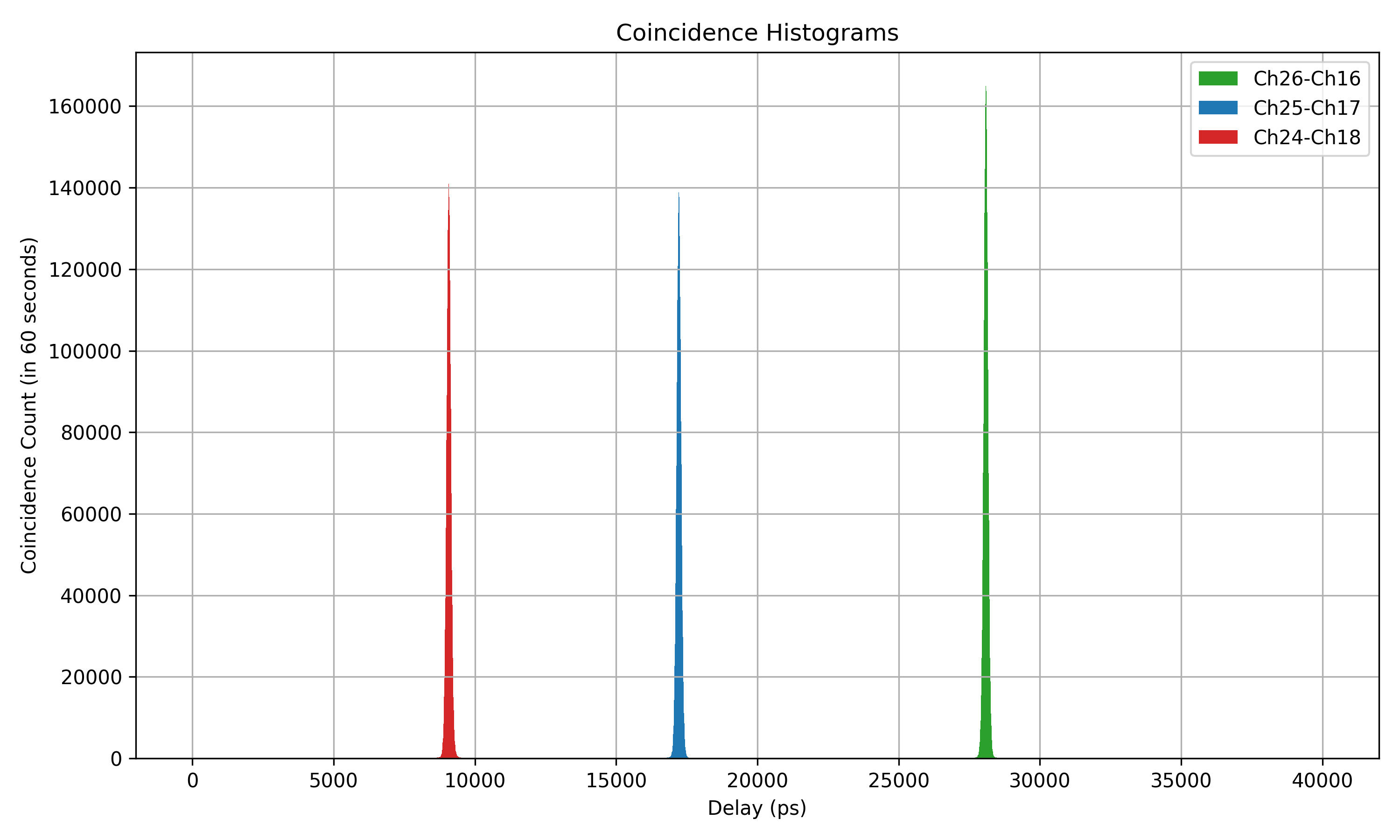}
\caption{Coincidence histograms for the three entangled channel pairs.}
\label{fig:coincidence_hist}
\end{figure}

Figure~\ref{fig:coincidence_hist} shows the coincidence histograms for the three entangled channel pairs: Ch26-Ch16, Ch25-Ch17, and Ch24-Ch18. Each histogram exhibits a distinct peak within a narrow time delay window, confirming strong temporal correlations between the signal and idler photons. The peaks are well-isolated and consistent across pairs, demonstrating reliable timing alignment and low background noise. These results validate the entanglement quality and timing precision of our measurement system over a 60-second acquisition interval.

\begin{figure}[tbp]
\includegraphics[width=\linewidth]{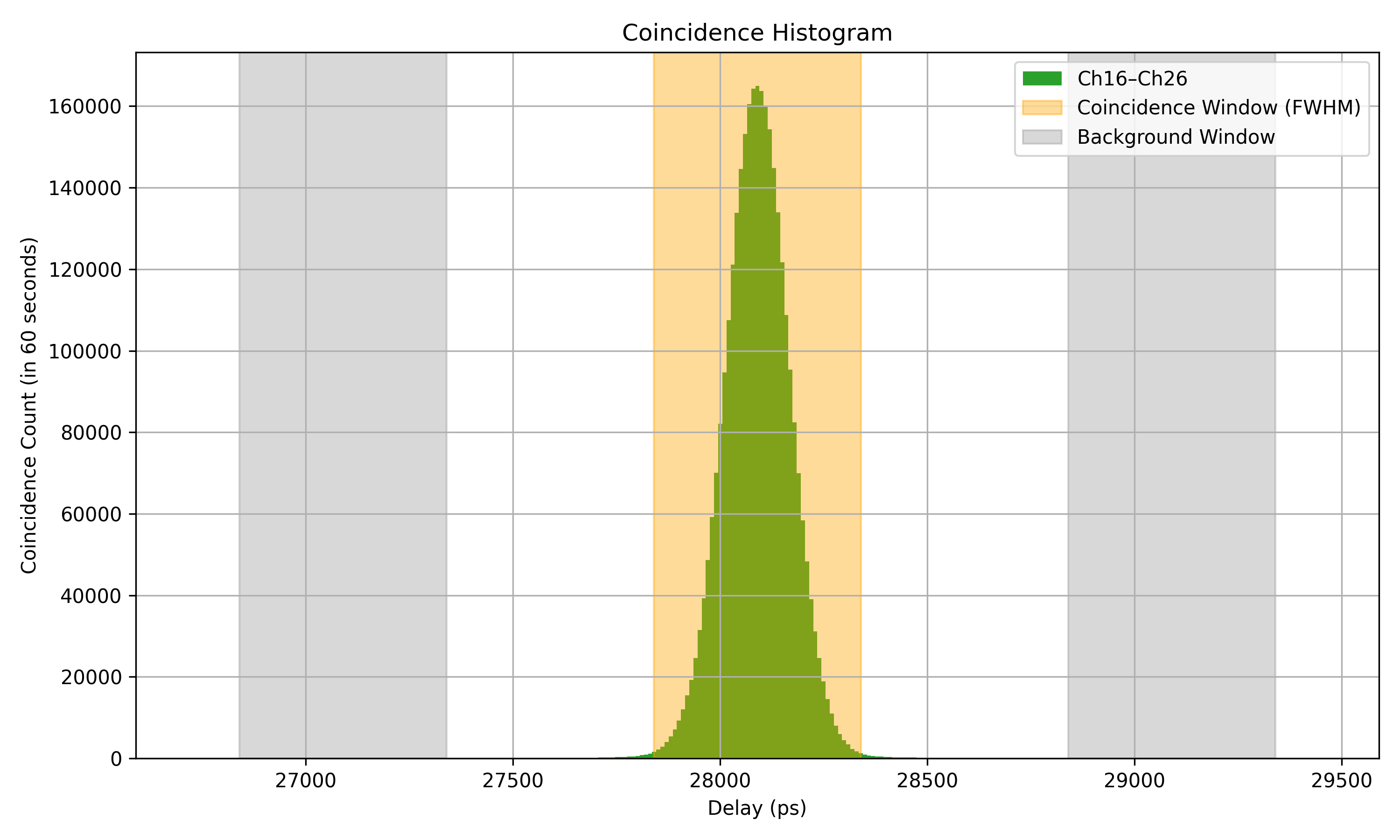}
\caption{Coincidence histogram for Ch26--Ch16 with a 500\,ps coincidence window (orange) and background windows (gray) placed 1\,ns from the peak, used for CAR estimation.}
\label{fig:coincidence_hist_windows}
\end{figure}

The coincidence histogram of Ch26--Ch16, shown in Figure~\ref{fig:coincidence_hist_windows}, illustrates the method used to define the coincidence and background windows for calculating CAR across three detector pairs. The coincidence count ($C_c$) is obtained by integrating the number of events within a 500\,ps window centered on the peak of the histogram. To estimate the accidental coincidence count ($A_{cc}$), two background windows of equal width (500\,ps) are positioned symmetrically 1\,ns away from the peak center, and the average count from these regions is used. The resulting CAR values for all three detector pairs are summarized in Table~\ref{tab:car_values}.

\begin{table}[tbp]
\caption{Coincidence and accidental count rates ($C_c$ and $A_{cc}$, respectively), along with CAR values for each detector pair.}
\label{tab:car_values}
\begin{center}
\begin{tabular}{| c | c | c | c |}
\hline
\textbf{Detector Pair} & \textbf{$C_c$ (counts/s)} & \textbf{$A_{cc}$ (counts/s)} & \textbf{CAR} \\
\hline
\rule{0pt}{3ex} Ch26--Ch16 & 53106.45 & 33.35 & 1592.40 \\
\rule{0pt}{3ex} Ch25--Ch17 & 45601.10 & 28.80 & 1583.37 \\
\rule{0pt}{3ex} Ch24--Ch18 & 45738.53 & 31.03 & 1473.85 \\
\hline
\end{tabular}
\end{center}
\end{table}

%--------------------------------------SIMULATION-----------------------------------%

\subsection{Software Performance}
The algorithm was evaluated through multiple simulations with varying parameters, including the number of resources and users. The users' arrival, inter-arrival and service times were randomly generated using Poisson process. This model is widely used, as many real-world traffic patterns, such as those in telecommunications or network queues, closely approximate or follow Poisson-like arrivals. The simulation script is written using SimPy, a Python library used to model and simulate real-world systems and processes. The simulation assumes that users arrive on average every 10 units of time, with a mean service time of 60 units. The total simulation duration is set to 1,000 units of time. The coincidence rate for each channel pair is randomly selected between 18,000 and 68,000 counts per second. Since the algorithm depends on relative QoS across users rather than absolute coincidence rates, its performance remains unaffected even if the actual hardware values differ. This range reflects a target for potential future hardware improvements. Each simulation is repeated three times to ensure consistency and the results were averaged to obtain the final output.

\begin{figure}[t]
\includegraphics[width=\linewidth]{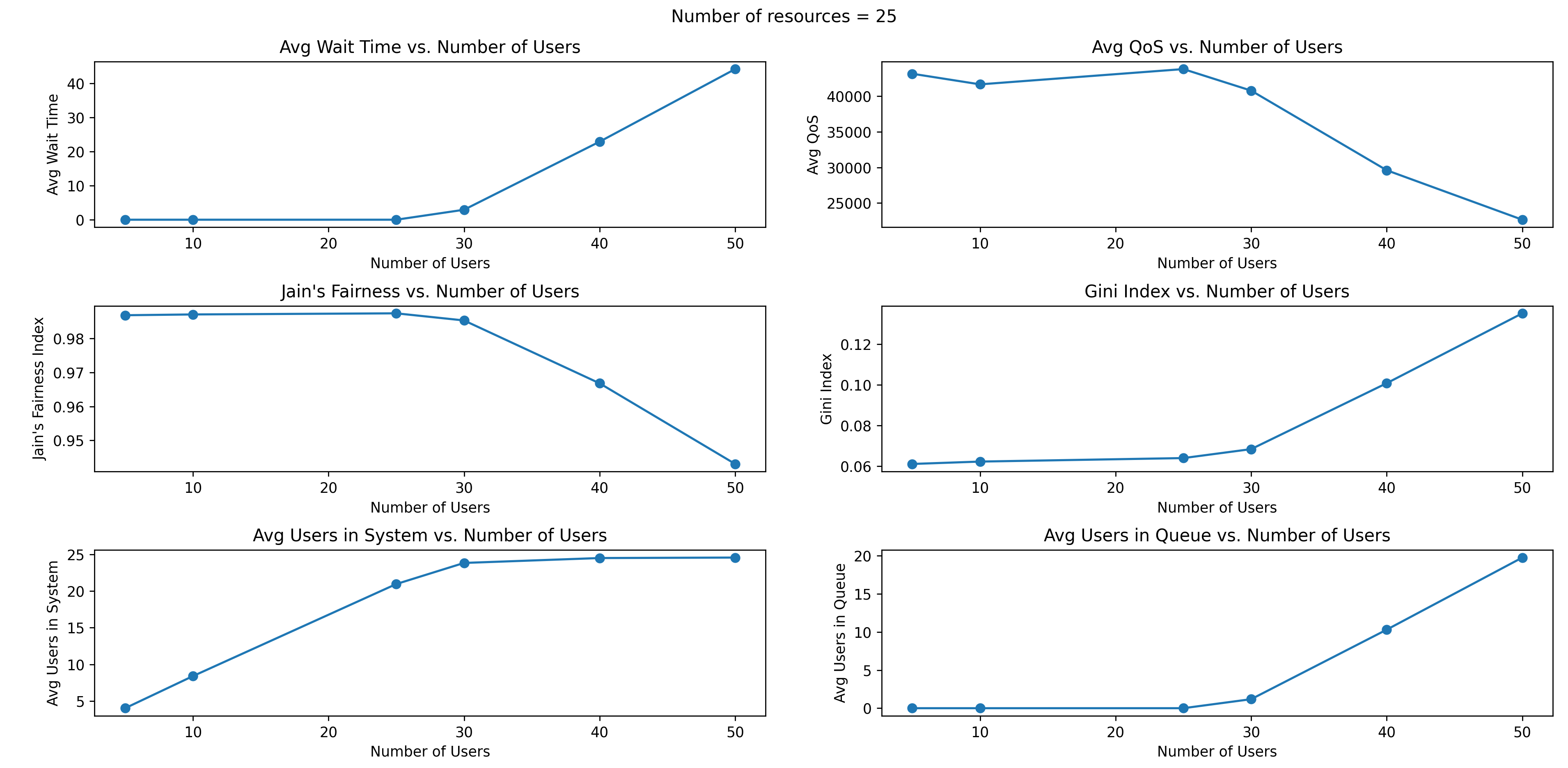}
\caption{System metrics vs.\ number of users for 25 resources.}
\label{fig:sim_vs_user}
\end{figure}

In Figure \ref{fig:sim_vs_user}, with 25 fixed resources, the system initially maintains relatively high QoS for moderate user loads. However, as the number of users rises beyond this capacity, average wait time and fairness metrics begin to degrade. This indicates that while the algorithm scales well within a certain range, it encounters saturation once too many users compete for the same finite set of channel links. Even then, the algorithm strives to preserve fairness, but its effectiveness is limited by the overall capacity constraints.

\begin{figure}[t]
\includegraphics[width=\linewidth]{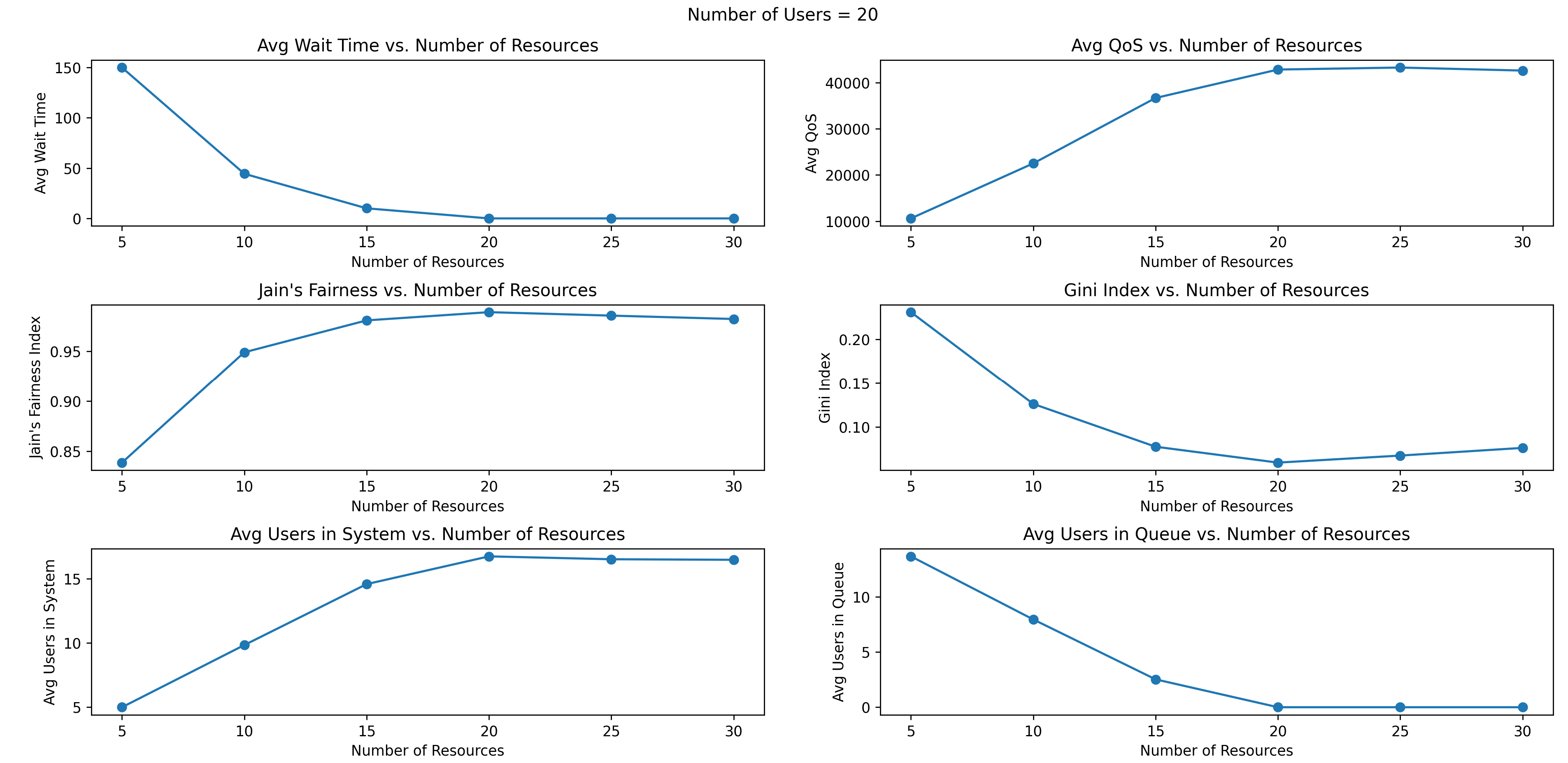}
\caption{System metrics vs.\ number of resources for 20 users.}
\label{fig:sim_vs_resource}
\end{figure}

In Figure \ref{fig:sim_vs_resource}, as the system adds more channel pairs, the resource allocation algorithm effectively reduces average wait times and increases QoS. Fairness also improves, suggesting that additional entanglement channels help the algorithm distribute resources more evenly among users. The resulting drop in queue length and wait times and high fairness metric at higher resource levels shows that the algorithm adapts by efficiently matching the same 20 users to a growing resource pool.

\begin{figure}[t]
\includegraphics[width=\linewidth]{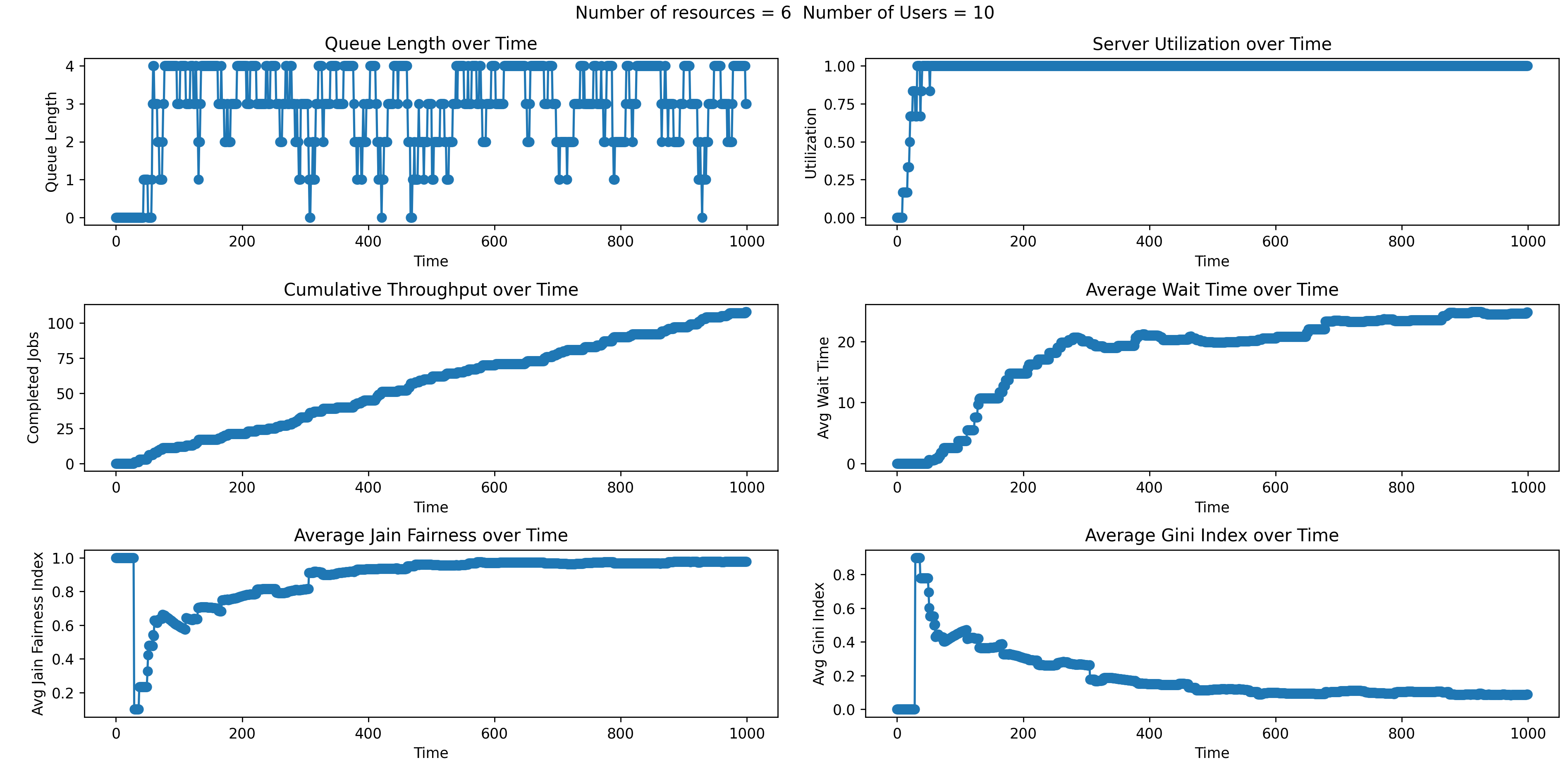}
\caption{Single simulation run results for 6 resources and 10 users.}
\label{fig:sim_single_run}
\end{figure}

For single simulation run with 6 resources and 10 users in the system, (Figure \ref{fig:sim_single_run}), the algorithm balances user arrivals and completions by oscillating the queue length between zero and four. Cumulative throughput grows steadily, and fairness metrics stabilize despite the system operating near capacity. Although average wait time naturally increases when resources are fully utilized due to queue backlog, the allocation strategy proves capable of adapting to changing demand. Over time, it continues to balance resource assignments such that no single user monopolizes the limited entanglement channels.

%-------------------Simulation Result vs User IMAGE --------------------%
%-------------------------------------------------------------------------%

%-------------------Simulation Result vs Resource IMAGE --------------------%
%-------------------------------------------------------------------------%

%-------------------Simulation Result single run IMAGE --------------------%
%-------------------------------------------------------------------------%

These results highlight the effectiveness of the resource allocation algorithm in optimizing fairness and reducing wait times while adapting to varying system conditions. Despite capacity constraints, the algorithm maintains a balanced distribution of resources, ensuring no single user dominates the available channels. As resource availability increases, it scales efficiently, demonstrating its ability to enhance overall system performance. These findings validate the Hungarian Matching algorithm’s ability to dynamically allocate quantum network resources while preserving fairness and maximizing QoS.

\subsection{Cloud Deployment and Testing}

We deployed the cloud application on Azure Kubernetes Service (AKS) for testing purpose. The local server - connected to the optical switch and the time tagger, was exposed to the public internet via Ngrok, which provides a secure tunnel without the need for additional network configuration. The test result showed that the cloud application and the local server were able to efficiently communicate with each other to process measurement results.

To evaluate the cloud application and simulate heavy load conditions, we conducted local load testing. We utilized Locust to generate concurrent API requests targeting the application’s backend, creating significant load. Simultaneously, the Kubernetes Horizontal Pod Autoscaler (HPA) dynamically adjusted the number of pods in response to CPU utilization. All load tests were performed on local machine, enabling us to validate both the autoscaling functionality and the application’s ability to withstand stress. For testing purpose the following each pod is initially allocated 35m CPU (equivalent to 0.035 vCPU), with a CPU usage limit set at 100m (0.1 vCPU). The system is configured to scale between a minimum of 1 replica and a maximum of 20 replicas based on demand. Horizontal pod autoscaling is triggered when the average CPU utilization exceeds 60\%. The number of concurrent users tested includes 20, 25, 50 and 100, with user interarrival times randomly selected between 1 and 5 units.

\begin{figure}[tbp]
\includegraphics[width=\linewidth]{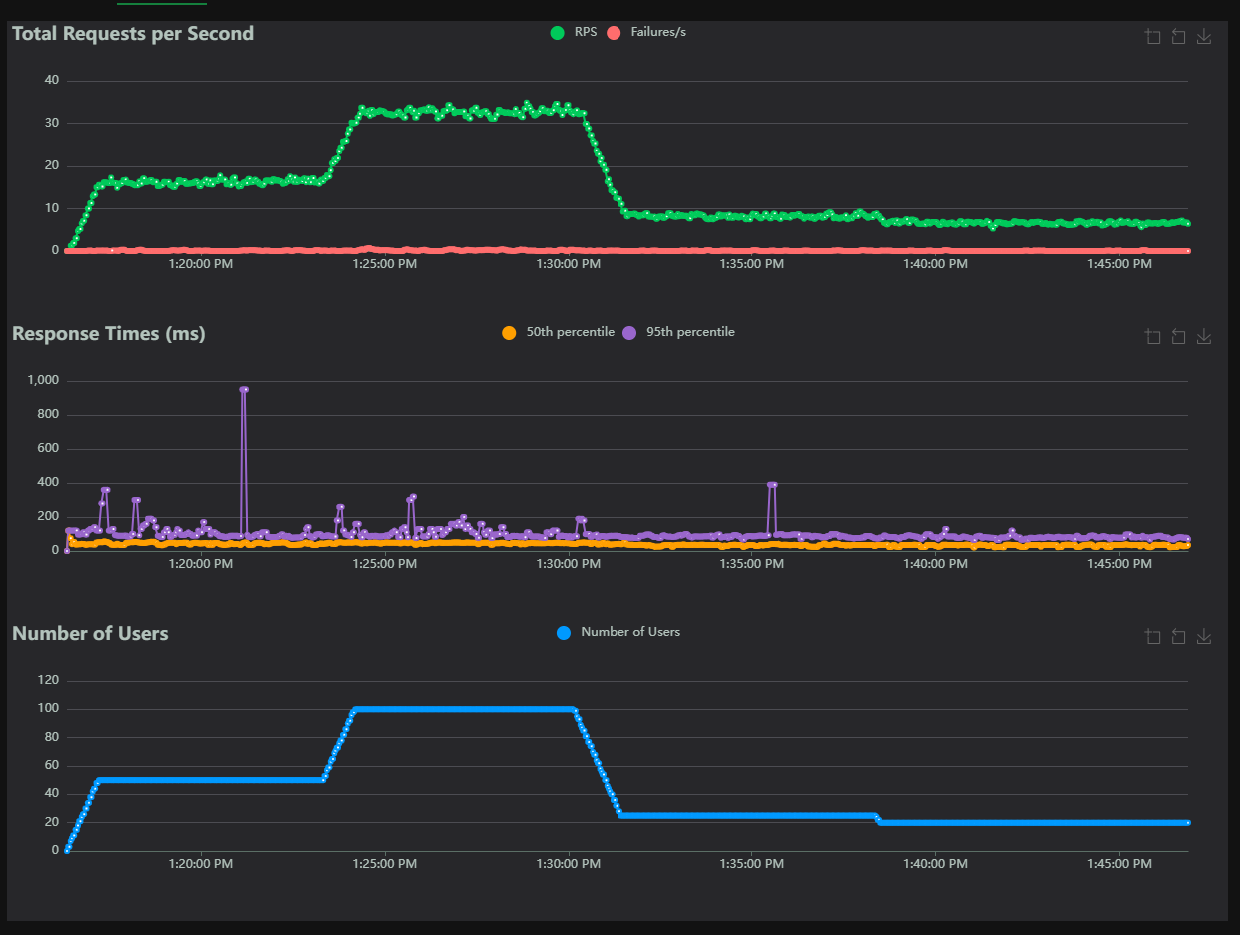}
\caption{Load testing dashboard illustrating request per second, latency and total number of users in the system. }
\label{fig:locust}
\end{figure}

Figure \ref{fig:locust} shows that as the number of users increases, the number of requests per second also rises, leading to higher latency and increased CPU utilization. In response, Kubernetes automatically scales up the number of running pods to handle the growing load, which in turn reduces latency. As the user count continues to rise, latency increases again but less sharply, as additional pods help absorb the demand, eventually reaching a peak of 17 pods. When the user load decreases, CPU utilization drops accordingly, and Kubernetes scales the pods down to 10, maintaining approximately 60\% CPU usage. Out of a total of 28,184 requests generated by Locust, only 139 requests failed, with an average user latency of 47.88 ms, highlighting Kubernetes’ ability to effectively adapt to changing workloads

%-------------------Locust graph IMAGE --------------------%
%-------------------------------------------------------------------------%

% %-------------------HPA IMAGE --------------------%
% \begin{figure}[tbp]
% \includegraphics[width=\linewidth]{figs/k8s_hpa_3.png}
% \caption{Scaling of pods replicas with respect to average CPU utilization.}
% \label{fig:hpa}
% \end{figure}
% %-------------------------------------------------------------------------%

% %-------------------------Locust Stats IMAGE --------------------------%
% \begin{figure}[tbp]
% \includegraphics[width=\linewidth]{figs/locust_stats_3.png}
% \caption{Aggregated metrics for all the request sent to the cloud application by Locust.}
% \label{fig:locust_stats}
% \end{figure}
% %-------------------------------------------------------------------------%

%-------------------------------------CONCLUSION-----------------------------------%

\section{Conclusion and Future Work}
This research demonstrates a novel approach to democratizing a quantum network testbed through software virtualization and fair resource allocation. By abstracting the complexities of quantum hardware, our system enables open access and experiment with quantum resources. The integration of a cloud-based architecture with a virtualized time tagger and optical switch allows multiple users to perform quantum measurements concurrently. Our resource allocation algorithm, which incorporates a proportional fair utility function and the Hungarian matching algorithm, has proven effective in balancing the distribution of limited entangled photon channel links among users. %Simulation results indicate that as additional resources are introduced, overall system performance improves—manifested by reduced wait times, higher quality of service, and enhanced fairness. Moreover, our deployment on Azure Kubernetes Service with dynamic scaling under heavy load conditions confirms the scalability and resilience of our design, paving the way for more accessible quantum network experimentation.

Future research will focus on expanding and refining the system to address evolving hardware constraints and user demands. As quantum hardware capabilities advance, new constraints such as variable detector efficiencies and additional channel configurations will need to be integrated into the resource allocation framework. Enhancing backend logic to handle detailed user session management, including tracking entanglement counts, usage times, and waiting periods, will provide deeper insights into system performance and user experience. %Additionally, developing advanced analytics and interactive dashboards will enable real-time visualization of key metrics such as entanglement quality, latency, and throughput, thereby facilitating better system monitoring and optimization. We also plan to explore more sophisticated scheduling techniques, including reservation-based and hybrid scheduling models, to further improve fairness and resource utilization under peak load conditions. These improvements will not only make the platform more robust and user-friendly but also foster greater innovation in quantum networking research.

%----------------------------------END--------------------------------%

\section*{Acknowledgment}
The work was supported in part by the National Science Foundation under grants 2304118 and 2326746.

% \sloppy
\bibliographystyle{IEEEtran}
\bibliography{citations}

\end{document}